\documentclass[letterpaper]{article}
\usepackage[top=3cm,bottom=3cm,left=3cm,right=3cm,marginparwidth=1.75cm]{geometry}
\usepackage{listings}
\usepackage{xcolor}
\usepackage{multirow}
\usepackage{dcolumn}
\usepackage[english]{babel}
\usepackage[utf8x]{inputenc}
\usepackage[T1]{fontenc}
\usepackage{palatino}
\usepackage{amsmath}
\usepackage{afterpage}
\usepackage{graphicx, subcaption}
\usepackage[colorinlistoftodos]{todonotes}
\usepackage[colorlinks=true, linkcolor=blue, citecolor=blue]{hyperref}
\usepackage{url}
\usepackage{xspace}

\newcommand*{\lint}{\ensuremath{\lambda_\text{INT}}\xspace}
\newcommand*{\GeV}{\ensuremath{\text{Ge\kern -0.1em V}}}
\newcommand*{\promom}{\ensuremath{400\,\GeV/c}\xspace}
\newcommand*{\onefigwidth}{0.8\textwidth}
\newcommand{\tit}[1]{\noindent \textbf{#1}\\[1ex] \noindent}

\begin{document}

\vspace*{1.2cm}
\thispagestyle{empty}
\begin{center}
{\LARGE \bf The SHiP experiment at CERN}
\par\vspace*{7mm}\par
{\bigskip \large \bf Markus Cristinziani}

\bigskip

{\large \bf  e-mail: cristinz@uni-bonn.de}

\bigskip

{Physikalisches Institut, Universit\"at Bonn, Nussallee 12, 53115 Bonn, Germany}

\bigskip

{\it Presented at the 3rd World Summit on Exploring the Dark Side of the Universe \\Guadeloupe Islands, March 9-13 2020}
\end{center}

\begin{abstract}
The current status of the proposed SHiP experiment at the CERN Beam Dump
Facility is presented.  SHiP is a general-purpose fixed-target experiment. The
\promom  proton beam extracted from the SPS will be dumped on a heavy target to
integrate $2 \times 10^{20}$ protons on target in five  years. The detector,
based on a long vacuum tank followed by a spectrometer  and particle
identification  detectors, will allow to probe a variety of  models with light
long-lived exotic  particles and masses below ${\cal  O}(10)\,\GeV /c^2$.  The
main focus will be  the physics of the so-called  hidden portals, i.e.\ the
search for dark photons,  light scalars and  pseudo-scalars, and heavy
neutrinos.  The sensitivity to  heavy neutrinos will allow to probe, in the
mass range  between the kaon and the charm meson mass,  a coupling range for
which  baryogenesis and active neutrino masses could also  be explained.  A
second  dedicated detector will study neutrinos and explore  light dark matter.
\end{abstract}

\section{Introduction}
SHiP is a general-purpose beam-dump experiment designed to use the
high-intensity \promom beam of protons available from the CERN SPS accelerator
in order to search for hidden particles. The proposal of the experiment evolved
from a first idea~\cite{Bonivento:2013jag} to search for Heavy Neutral Leptons
(HNLs), whose existence is predicted in the $\nu$MSM model~\cite{Asaka:2005an,
Asaka:2005pn} with guidance described in Ref.~\cite{Gorbunov:2007ak}, to a more
comprehensive programme of new physics searches at the intensity
frontier~\cite{Alekhin:2015byh}, including the option of producing and studying
a large sample of $\nu_\tau$ neutrinos~\cite{DeLellis:2015usa}, that lead to
the SHiP technical proposal~\cite{Anelli:2015pba}. Since then, the detector
layout has been continuously optimised and expanded. The most recent results
were documented in a progress report~\cite{Ahdida:2654870} and the
comprehensive design study report~\cite{Ahdida:2704147}.

\section{Motivation}
While the Higgs boson discovery at the Large Hadron Collider in 2012 marks the
triumph of the Standard Model (SM), there are still several shortcomings in
particle physics that are waiting to be explained. The SM particles cannot
account for the observed matter in the Universe, as there is convincing
evidence of dark matter (DM), with unknown mass and couplings. The predominance
of matter over antimatter in the Universe calls for additional sources of $CP$
violation, beyond what is known in the SM. The tiny, but non-zero masses of the
neutrinos, causing oscillations, could be explained via the see-saw mechanism
with right-handed neutrinos with Yukawa couplings to the Higgs boson and SM
leptons. There are also aesthetic shortcomings, like the strong $CP$ problem or
the hierarchy/fine-tuning problem of the Higgs mass.

It is thus important to probe new physics beyond the SM in different
directions.  At the intensity frontier the paradigm is that low-energy,
high-intensity experiments could uncover new interactions and particles with
very feeble couplings. Popular extensions of the SM proceed through so-called
portals, that are generally categorized according to the nature of the
mediator: scalar, vector, pseudo-scalar or fermion. Examples for these
categories are Dark Higgs, Dark Photon, Axion Like Particles (ALPs), and HNLs,
respectively.

\section{The SHiP detector at the Beam Dump Facility}
\begin{figure}[htbp]
\centering
\includegraphics[width=\onefigwidth]{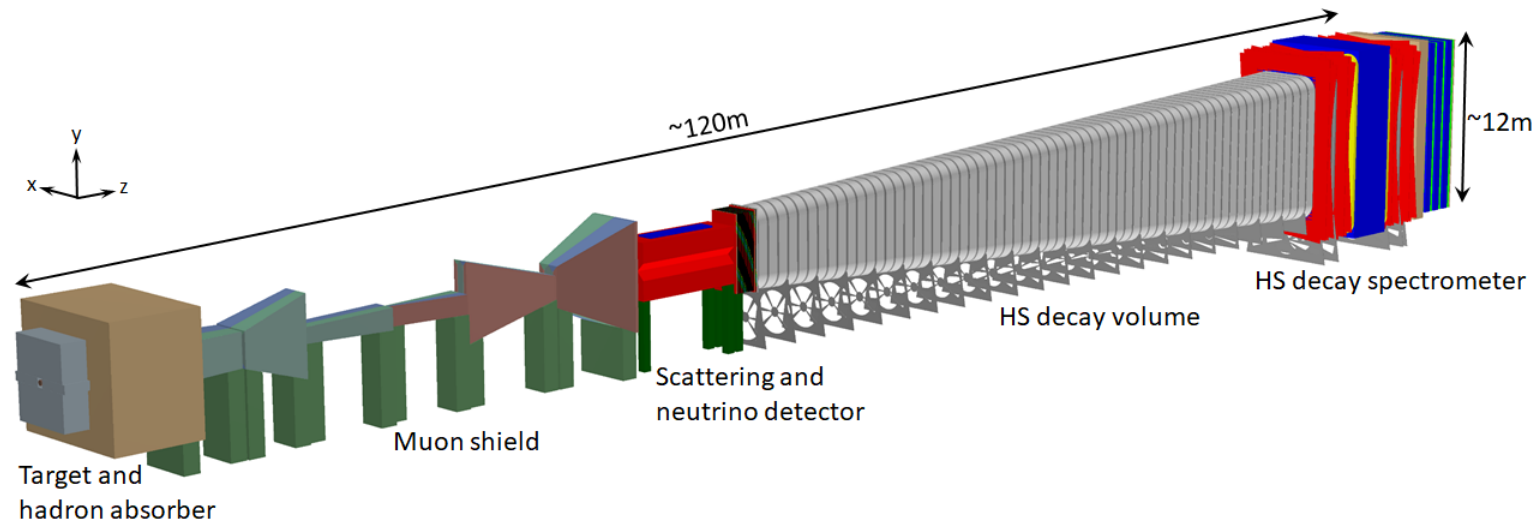}
\caption{\label{fig:detector} Drawing of the current outline of the
SHiP detector~\cite{Ahdida:2704147}.}
\end{figure}

The SHiP detector is planned to be installed in the North Area at the CERN SPS
400\,\GeV/$c$ beam, taking advantage of the Beam Dump Facility (BDF). It
features two main sub-detector systems, pursuing complementary physics goals.
The hidden sector (HS) decay spectrometer aims at measuring visible decays of
HS particles and will reconstruct their decay vertices in a large decay volume.
The scattering and neutrino detector (SND) is dedicated to neutrino physics and
the search for light dark matter (LDM).  The design of the detector underwent
several optimisation steps and its current implementation is shown in
Figure~\ref{fig:detector}. In particular, there has been a large effort to
re-optimize the muon magnetic shield configuration and, as a consequence, the
detector layout.

\subsection{Beam Dump Facility}
The proposed BDF is foreseen to be located at the North Area of the CERN SPS.
It is designed to be able to serve both, beam-dump like and fixed-target
experiments.  The full exploitation of the SPS would allow the delivery of an
annual yield of up to $4 \times 10^{19}$ 400\,\GeV/$c$ protons on target (pot)
while delivering protons to the HL-LHC and the existing SPS facilities. Slow
extraction of a de-bunched beam with good uniformity is required to keep the
combinatorial background under control and to dilute the large beam power in
the proton target. The proposed implementation of the beam dump experimental
facility is based on a minimal modification to the SPS complex and maximal
use of the existing beamlines. The BDF is described in detail in
Ref.~\cite{Ahdida:2019ubf}.

\subsection{Target}
The SHiP target~\cite{LopezSola:2019sfp} is designed to maximise the production
of hidden sector particles, mainly through the decay of flavoured hadrons
produced in the target through primary and secondary (cascade) interactions. At
the same time, the target material and density are chosen such to reduce the
production of other particles. The longitudinally segmented hybrid target
consists of four nuclear interaction lengths (\lint) of TZM (titanium-zirconium
doped molybdenum) alloy, followed by 6\lint of tungsten, and is interleaved
with water cooling spaces, for a cross section of $30 \times 30\,\text{cm}^2$
and a total length of more than 120\,cm. Cooling will be essential, as the peak
power of the beam during a spill will be 2.56\,MW. The target complex will be
housed in a 440\,m$^3$ bunker made of remotely handled iron bricks and is
additionally cooled, underpressurized, and shielded. A 5\,m thick iron shield
absorbs hadrons behind the target.

\subsection{Active muon shield}
After the hadron absorber, penetrating muons are a possible source of undesired
occupancy. From simulation, it is expected that $10^{11}$\,muons/s are produced
in the collisions of the proton beam on the beam-dump target from the decay of
pions, kaons, and other light and charmed mesons. A magnetic muon
shield~\cite{Akmete:2017bpl} will deflect muons of both polarities and a wide
range of transverse momenta and significantly reduce this muon flux. Since the
produced hidden sector particles typically exhibit large transverse momenta,
the muon shield has to be as compact as possible. The original design has been
revised using a Bayesian optimisation algorithm~\cite{Baranov:2017chy},
resulting in a total length of 34\,m, a weight of 1.5\,kt, and a magnetic field
of up to 1.8\,T. The first part aims at separating muons of opposite polarities
and also acts as a hadron absorber, while the rest of the series of magnets is
designed to also remove the lower momentum muons that re-enter the acceptance
due to the return fields. The muon flux will be reduced to $10^5$\,muons/s
after the muon shield.

\subsection{Scattering and neutrino detector}
Downstream of the muon shield a detector system consisting of emulsion cloud
chamber (ECC) bricks made of lead plates and nuclear emulsion films,
interleaved with electronic trackers and followed by a compact emulsion
spectrometer (CES) with low-density material, will be immersed in a 1.2\,T
magnetic field. Behind the magnet, a muon identification system, consisting of
several 10\,cm thick iron filters and RPC tracking planes, aims at identifying
the muons produced in neutrino interactions and $\tau$-decays occurring in
the emulsion target.  The ECC bricks are composed of stacks of alternating
1\,mm lead plates, acting as neutrino targets, and 300\,$\mu$m emulsion films,
where $\nu_\tau$ interactions and $\tau$-lepton decay vertices can be
reconstructed. The CES, a light structure with a long lever arm, will be
essential to measure the charge of the $\tau$-lepton daughters.  The electronic
tracker technologies under study include scintillating fibres (SciFi), and
micro-pattern gaseous detectors ($\mu$-RWELL).

\subsection{Hidden sector decay volume}
The SHiP decay volume has a pyramidal frustum shape with rectangular bases of
2.4 $\times$ 4.5\,m$^2$ at the entrance and 5 $\times$ 10\,m$^2$ at the exit,
for a length of 50\,m. It has to be sufficiently evacuated from residual air
to suppress muon and neutrino interactions.  The complete volume is
enclosed by the surrounding background tagger (SBT), a veto detector, based on
480\,t of linear alkylbenzene liquid scintillators, which will help to identify
activity from outside of the detector, like cosmic rays or cavern background.
The liquid scintillator is coupled to 3\,500 wavelength-shifting optical
modules (WOM), already proposed to be used in an extension of the IceCube
detector.

\subsection{Spectrometer and particle identification}
The HS decay volume is followed by a spectrometer with an acceptance of 5
$\times$ 10\,m$^2$. The first sub-detector is the spectrometer straw tracker,
designed to accurately reconstruct the momenta and the decay vertex, mass and
impact parameter of the hidden particle trajectory at the proton target. The
spectrometer dipole magnet is a conventional magnet with a total field integral
of 0.65\,Tm.  The SplitCal electromagnetic calorimeter employs the sampling
technology with lead/scintillator planes.  To accurately measure the shower
transverse profile, the SplitCal is longitudinally segmented and will contain
high-precision layers of MicroMegas chambers, similar to those developed for
the ATLAS muon upgrade.  This will be important for the reconstruction of
decays of axion-like particles to two photons. The muon system consists of four
stations interleaved by three muon filters.  To reject random
crossings, a dedicated timing detector made of scintillator bars is placed at
the end of the detector and will match the arrival times of the particles
forming vertex candidates with a precision of 100\,ps.

\section{Physics performance}

The SHiP experiment is a unique discovery tool for HS particles. Present
constraints on various channels will be improved by several orders of
magnitude. SHiP will also distinguish among different models, and measure
parameters relevant for cosmology and model building, in a large part of the
parameter space. Together with the direct search for LDM and neutrino physics,
SHiP is a wide scope general-purpose fixed-target experiment.

The HS detector is designed to fully reconstruct a wide range of decay modes,
and identify the particles, to ensure a model-independent search for
hidden particles.  The SND detector is optimised for scattering signatures of
LDM and neutrinos.  The sensitivity of the SHiP experiment heavily relies on
redundant background suppression.  The SHiP physics case was presented in
Ref.~\cite{Alekhin:2015byh}.

\subsection{Background studies}
Large samples of simulated events have been produced to accurately determined
the level of backgrounds, including from the floor, ceiling, walls and detector
supports, using Geant4~\cite{Geant} and the FairRoot framework~\cite{FairRoot}.
Samples of muons produced in $10^{12}$\,pot have been fully simulated. In order
to claim discovery of a HS particle, it is of paramount importance to suppress
the backgrounds that would mimic the same final states to a negligible level.
Signal events feature a vertex pointing back to the target, consisting of at
least two charged particles and possibly additional invisible particles.

Three sources of background can mimic the HS signature: random muon
combinatorial, muon inelastic scattering and neutrino deep-inelastic
scattering. The background from cosmics has been found to be negligible. The
three sources of background are reduced by  requirements on the track momentum,
vertex position and impact parameter with respect to the target. To
avoid irreducible background from neutrinos interacting with the air molecules
inside the vessel, a level of vacuum below $10^{-2}$\;bar is necessary. The
background from neutrino scattering in the floor and the walls of the cavern
was studied and found to be negligible. Additionally, information from the
timing detector and veto information from the SBT will reduce the backgrounds
to less than $0.3$ events during the five-year data-taking. Thus, SHiP can be
considered a zero-background experiment with a high level of redundancy to
efficiently suppress the background for a broad spectrum of searches.

\subsection{Signal sensitivities}

\begin{figure}[htbp]
\begin{center}
\includegraphics[width=0.543\textwidth]{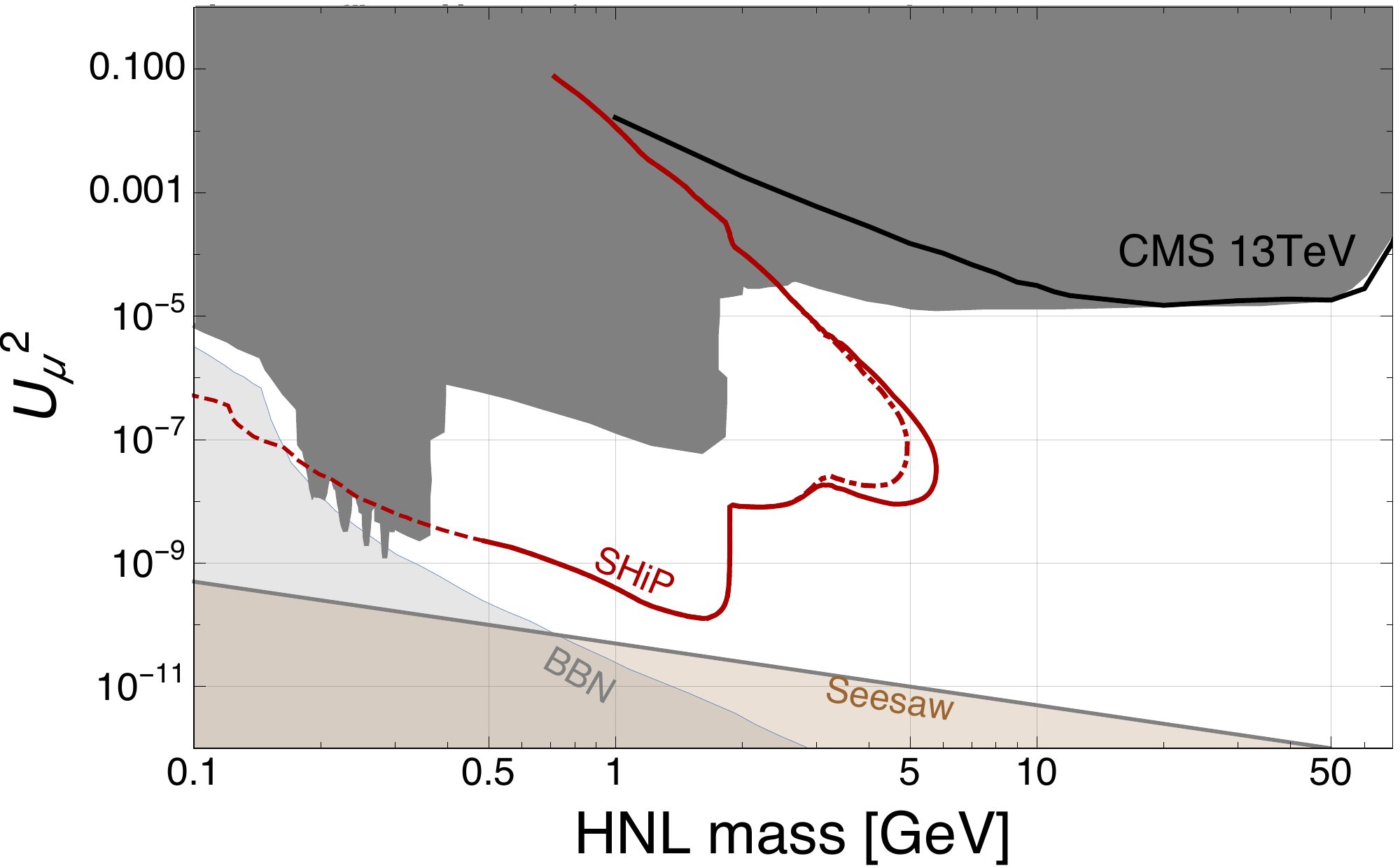}
\includegraphics[width=0.443\textwidth]{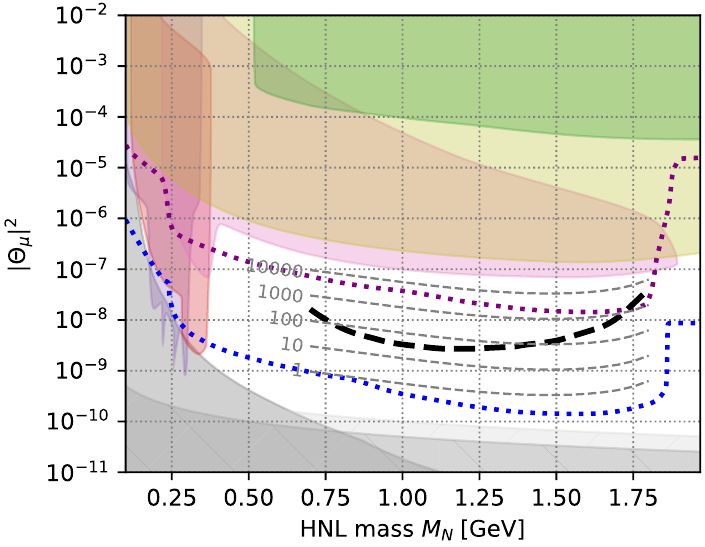}
\caption{Left: Sensitivity to HNL for the mixing with muon
flavour~\cite{SHiP:2018xqw}. The dark grey area and the solid line indicate the
excluded regions by previous experiments. The solid and dashed-dotted red lines
indicate the uncertainty related to the production of $B_c$ mesons. Right: SHiP
sensitivity to lepton number violation (thick dashed curve) compared to
exclusions by previous experiments (coloured areas).  The thin dashed grey
lines show the number of fully reconstructible events~\cite{Tastet:2019nqj}.
\label{fig:HNL}}
\end{center}
\end{figure}

\begin{figure}[htbp]
\begin{center}
\includegraphics[width=0.46\textwidth]{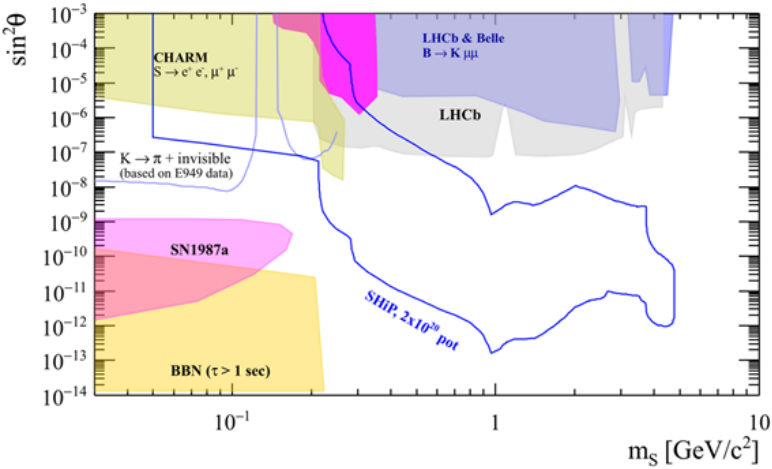}
\includegraphics[width=0.49\textwidth]{fig03b}\\[2mm]
\includegraphics[width=0.49\textwidth]{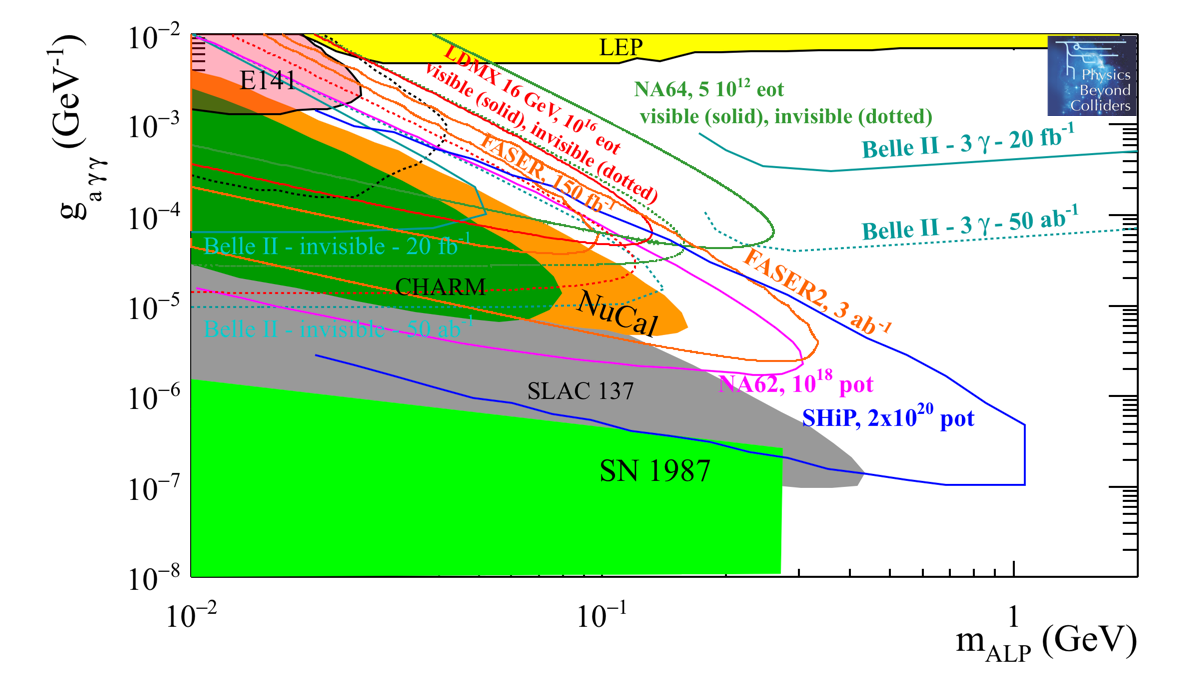}\hfill
\includegraphics[width=0.41\textwidth]{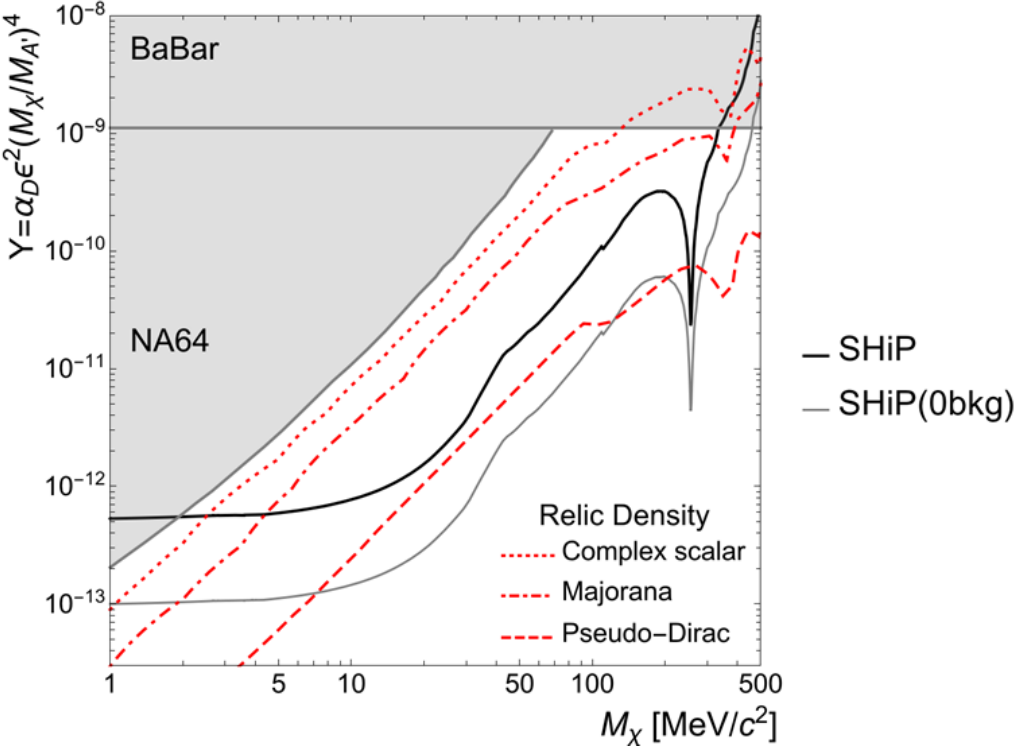}
\caption{\label{fig:dark} Sensitivity of SHiP to dark
scalar~\cite{Ahdida:2704147}, dark photon, ALPs~\cite{Beacham:2019nyx} and LDM
signals~\cite{Ahdida:2704147}, compared to existing (coloured or grey areas)
and projected (lines) exclusion limits.  }
\end{center}
\end{figure}

\tit{Hidden sector}
To illustrate the physics potential of SHiP to hidden sector particles the
sensitivities to HNLs, dark scalars, dark photons and ALPs coupling to photons
are shown in Figures~\ref{fig:HNL} and \ref{fig:dark}.  More details can be
found in the report of the Physics Beyond Colliders study
group~\cite{Beacham:2019nyx}.

In the case of a discovery, SHiP is capable of measuring parameters and identifying
the underlying models. For instance, SHiP may distinguish between Majorana-type
and Dirac-type HNLs in a significant fraction of the parameter space by
detecting lepton number violating or conserving decays~\cite{Tastet:2019nqj},
as shown in Figure~\ref{fig:HNL}.  If the mass splitting between the HNLs is
small, it may also be possible to resolve HNL oscillations as a direct
measurement of the mass splitting between HNLs.

\clearpage

\tit{Light Dark Matter}
If LDM particles undergo elastic scattering ($\chi e^- \rightarrow \chi e^-$)
in the SND detector material, the electromagnetic shower induced by the recoil
electron can be detected in the SND, which would act as a sampling calorimeter.
A sufficient portion of the shower can be reconstructed in order to determine
the particle angle and energy. The high accuracy of the nuclear emulsions will
provide topological discrimination against neutrino-induced background.
Figure~\ref{fig:dark} shows the SHiP sensitivity as a function fo the LDM mass
$M_\chi$, along with existing constraints and the thermal relic abundance, for
a benchmark scenario with a dark coupling $\alpha_D = 0.1$.\\[0ex]

\tit{Neutrino physics}
The SND detector will be able to determine the neutrino flavour by measuring
the flavour of the charged lepton produced in the neutrino charged-current
interactions. About $10^4$ $\tau$-neutrinos will be detected and the tracking
capabilities in the SND will allow distinguishing, for the first time, between
$\nu_\tau$ and $\bar{\nu_\tau}$.  In addition, the large sample of
neutrino-induced charm production will allow for unprecedented studies in this
domain, as for instance double-charmed production or the strange-quark content
of nucleons.  The samples available at SHiP will also allow to significantly
constrain the $\nu_\tau$ magnetic moment and test lepton flavour violation in
the neutrino sector.

\section{Measurements}
\subsection{Muon flux normalisation}

In order to validate the Monte-Carlo simulation (Pythia, Geant4) that is  being
used for the sensitivity studies, the muon flux was measured in a test-beam
setup at the  CERN SPS~\cite{Ahdida:2020doo}.  Protons of $400\,\GeV/c$ were
directed onto a SHiP target replica with full length. Behind an iron hadron
absorber the emanated muons were measured with a spectrometer setup using
scintillators, OPERA drift-tubes stations, the Goliath magnet and an RPC-based
muon tagger.  In three weeks during 2018, about 327 billion pot were recorded,
corresponding to $1\%$ of a nominal SHiP spill.  Events with muons were
recorded at a rate of one in 710.

\begin{figure}[htbp]
\begin{center}
\includegraphics[width=0.480\textwidth]{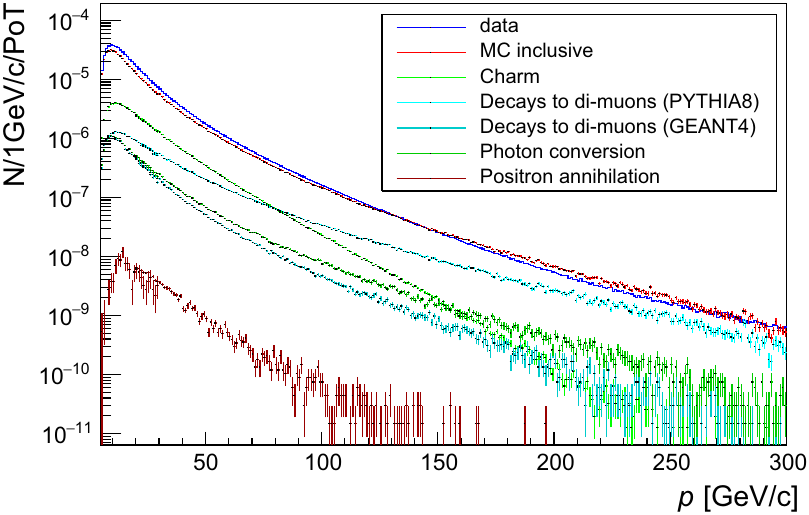}
\raisebox{3ex}{\includegraphics[width=0.506\textwidth]{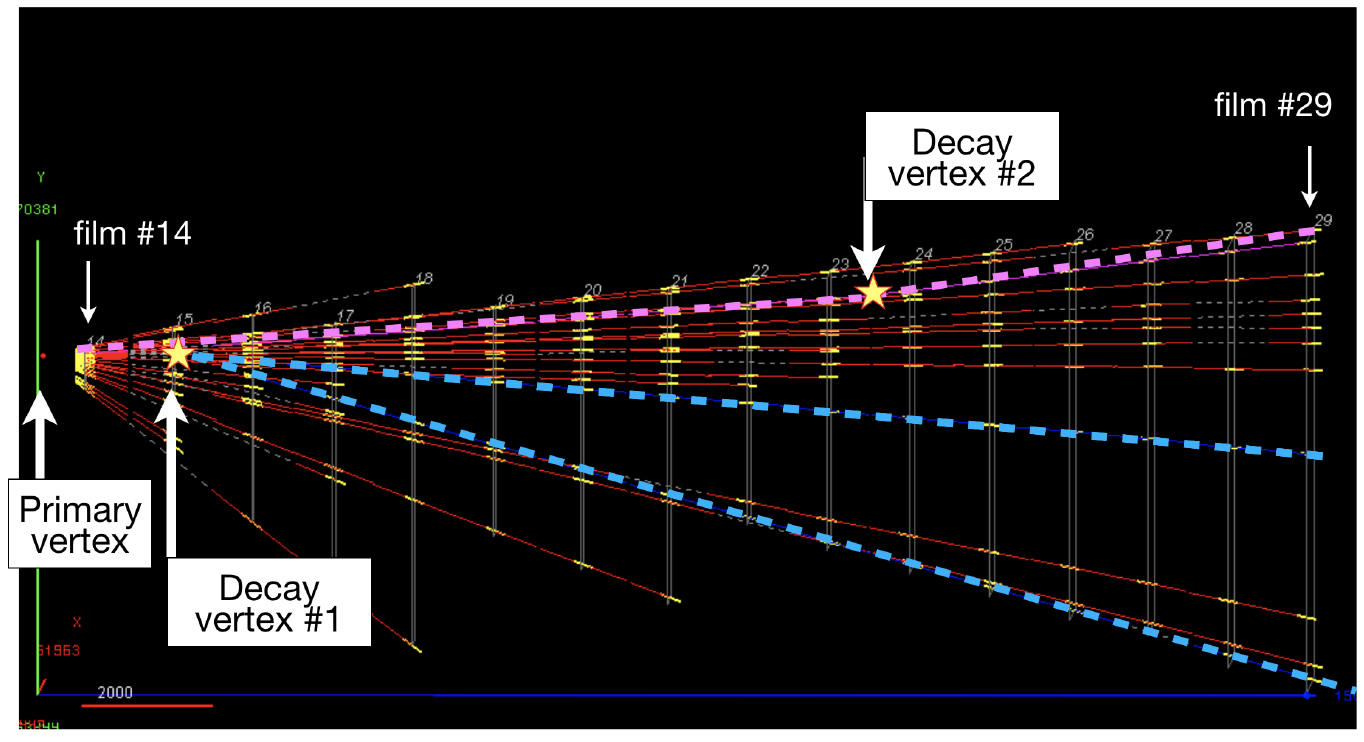}}
\caption{Left: Measured muon momentum distribution compared to
simulation~\cite{Ahdida:2020doo}.  The most important individual sources in
simulation are shown, along with the total prediction. Right: A double-charm
candidate event produced in proton collisions with the tungsten target recorded
at the H4 charm-production measurement~\cite{Ahdida:2704147}.
\label{fig:testbeams}}
\end{center}
\end{figure}

The relevant physics processes for muon production are foremost the decays of
pions and kaons, the production and decay of charm particles and low-mass
resonances, and the transportation of the muons through the iron absorber.
Data and simulation agree remarkably well (Figure~\ref{fig:testbeams}), with
maximal differences in the absolute rate of $30\%$ for large transverse momenta
at high muon momenta.

\subsection{Charm production cross-section}
The production of charmed hadrons in $400\,\GeV/c$ protons on the SHiP target
is expected to be increased by a factor of more than two with respect to the
direct production due to the interactions of particles produced in the hadronic
cascade showers. The exact normalisation is an important input and will be
measured by the SHiP-charm experiment at the CERN SPS~\cite{Akmete:2286844}.

An optimisation run was performed in 2018, collecting about 1.5 million pot,
directly after the muon flux measurement, thus using the same magnet, the
scintillators, drift tubes and RPCs. To precisely measure the charm production
vertices and the tracks, emulsion films were used, along with a tracker built
with ATLAS pixel modules and LHCb SciFi. To keep the occupancy to
manageable levels, the emulsion detector was moving horizontally during a spill
at $2$ cm/s and vertically in between spills. This makes the alignment and
matching of tracks between the detectors challenging. Preliminary results show
that more than $50\%$ of the tracks associated with reconstructed vertices can be
matched. Several runs were taken with varying amounts of target material in
front of the emulsion detector, to probe the charm content of the
different parts of the shower development. Figure~\ref{fig:testbeams} shows an
event display of a charm candidate in the emulsion detector.

\section{Conclusions}

With the current status of SHiP and the mature understanding of the continued
detector developments, the SHiP project is ready to commence the Technical
Design Report phase. First prototypes of all subsystems can be constructed and
tested within the next three years. It is estimated that the detector
production will require two to three years and that the detector assembly and
installation, including infrastructure, will require another two years. The
installation of the facility and the SHiP detector could be performed in the
Long Shutdown 3, allowing commissioning and starting data-taking early in Run
4.

\section*{Acknowledgements}

The author would like to thank the organisers of the 3rd World Summit on
Exploring the Dark Side of the Universe 2020 for a very enjoyable conference in
Guadeloupe and the Physique-Outremer association. The work of the author is
funded by the German Science Foundation (DFG) under Grant Agreement MC312/4-1.

\bibliographystyle{JHEP}
\bibliography{main}

\providecommand{\href}[2]{#2}\begingroup\raggedright\begin{thebibliography}{10}

\bibitem{Bonivento:2013jag}
W.~Bonivento et~al., {\it {Proposal to Search for Heavy Neutral Leptons at the
  SPS}},  \href{http://arxiv.org/abs/1310.1762}{{\tt arXiv:1310.1762}}.

\bibitem{Asaka:2005an}
T.~Asaka, S.~Blanchet, and M.~Shaposhnikov, {\it {The $\nu$MSM, dark matter and
  neutrino masses}},  {\em Phys. Lett.} {\bf B631} (2005) 151,
  [\href{http://arxiv.org/abs/hep-ph/0503065}{{\tt hep-ph/0503065}}].

\bibitem{Asaka:2005pn}
T.~Asaka and M.~Shaposhnikov, {\it {The $\nu$MSM, dark matter and baryon
  asymmetry of the universe}},  {\em Phys. Lett.} {\bf B620} (2005) 17,
  [\href{http://arxiv.org/abs/hep-ph/0505013}{{\tt hep-ph/0505013}}].

\bibitem{Gorbunov:2007ak}
D.~Gorbunov and M.~Shaposhnikov, {\it {How to find neutral leptons of the
  $\nu$MSM?}},  {\em JHEP} {\bf 10} (2007) 015,
  [\href{http://arxiv.org/abs/0705.1729}{{\tt arXiv:0705.1729}}].
  \hspace{-0.7ex}, Erratum and Addendum:
  \href{http://doi.org/10.1007/JHEP11(2013)101}{JHEP {\bf 11} (2013) 101}.

\bibitem{Alekhin:2015byh}
S.~Alekhin et~al., {\it {A facility to Search for Hidden Particles at the CERN
  SPS: the SHiP physics case}},  {\em Rept. Prog. Phys.} {\bf 79} (2016)
  124201, [\href{http://arxiv.org/abs/1504.04855}{{\tt arXiv:1504.04855}}].

\bibitem{DeLellis:2015usa}
G.~De~Lellis, {\it {Search for Hidden Particles (SHiP): a new experiment
  proposal}},  {\em Nucl. Part. Phys. Proc.} {\bf 263} (2015) 71.

\bibitem{Anelli:2015pba}
{\bf SHiP} Collaboration, M.~Anelli et~al., {\it {A facility to Search for
  Hidden Particles (SHiP) at the CERN SPS}},
  \href{http://arxiv.org/abs/1504.04956}{{\tt arXiv:1504.04956}}.

\bibitem{Ahdida:2654870}
{\bf SHiP} Collaboration, {\it {SHiP Experiment - Progress Report}},  Tech.
  Rep. CERN-SPSC-2019-010. SPSC-SR-248, CERN, Geneva, Jan, 2019.

\bibitem{Ahdida:2704147}
{\bf SHiP} Collaboration, {\it {SHiP Experiment - Comprehensive Design Study
  report}},  Tech. Rep. CERN-SPSC-2019-049; SPSC-SR-263, CERN, Geneva, Dec,
  2019.

\bibitem{Ahdida:2019ubf}
C.~Ahdida et~al., {\it {SPS Beam Dump Facility - Comprehensive Design Study}},
  \href{http://arxiv.org/abs/1912.06356}{{\tt arXiv:1912.06356}}.

\bibitem{LopezSola:2019sfp}
E.~Lopez~Sola et~al., {\it {Design of a high power production target for the
  Beam Dump Facility at CERN}},  {\em Phys. Rev. Accel. Beams} {\bf 22} (2019)
  113001, [\href{http://arxiv.org/abs/1904.03074}{{\tt arXiv:1904.03074}}].

\bibitem{Akmete:2017bpl}
{\bf SHiP} Collaboration, A.~Akmete et~al., {\it {The active muon shield in the
  SHiP experiment}},  {\em JINST} {\bf 12} (2017) P05011,
  [\href{http://arxiv.org/abs/1703.03612}{{\tt arXiv:1703.03612}}].

\bibitem{Baranov:2017chy}
A.~Baranov et~al., {\it {Optimising the active muon shield for the SHiP
  experiment at CERN}},  {\em J. Phys. Conf. Ser.} {\bf 934} (2017) 012050.

\bibitem{Geant}
{\bf GEANT4} Collaboration, S.~Agostinelli et~al., {\it {GEANT4 --- A
  Simulation toolkit}},  {\em Nucl. Instrum. Meth.} {\bf A506} (2003) 250.

\bibitem{FairRoot}
M.~Al-Turany et~al., {\it {The FairRoot framework}},  {\em J. Phys. Conf. Ser.}
  {\bf 396} (2012) 022001.

\bibitem{Beacham:2019nyx}
J.~Beacham et~al., {\it {Physics Beyond Colliders at CERN: Beyond the Standard
  Model Working Group Report}},  {\em J. Phys.} {\bf G47} (2020) 010501,
  [\href{http://arxiv.org/abs/1901.09966}{{\tt arXiv:1901.09966}}].

\bibitem{Tastet:2019nqj}
J.-L. Tastet and I.~Timiryasov, {\it {Dirac vs. Majorana HNLs (and their
  oscillations) at SHiP}},  {\em JHEP} {\bf 04} (2020) 005,
  [\href{http://arxiv.org/abs/1912.05520}{{\tt arXiv:1912.05520}}].

\bibitem{SHiP:2018xqw}
{\bf SHiP} Collaboration, C.~Ahdida et~al., {\it {Sensitivity of the SHiP
  experiment to Heavy Neutral Leptons}},  {\em JHEP} {\bf 04} (2019) 077,
  [\href{http://arxiv.org/abs/1811.00930}{{\tt arXiv:1811.00930}}].

\bibitem{Ahdida:2020doo}
{\bf SHiP} Collaboration, C.~Ahdida et~al., {\it {Measurement of the muon flux
  from 400 GeV/c protons interacting in a thick molybdenum/tungsten target}},
  {\em Eur. Phys. J.} {\bf C80} (2020) 284,
  [\href{http://arxiv.org/abs/2001.04784}{{\tt arXiv:2001.04784}}].

\bibitem{Akmete:2286844}
{\bf SHiP} Collaboration, {\it {Measurement of associated charm production
  induced by 400 GeV/c protons}},  Tech. Rep. CERN-SPSC-2017-033; SPSC-EOI-017,
  CERN, Geneva, Oct, 2017.

\end{thebibliography}\endgroup

\end{document}